\documentclass[11pt,superscriptaddress]{article}
\pdfoutput=1
\usepackage{graphicx}  % needed for figures
\usepackage{dcolumn}   % needed for some tables
\usepackage{bm}        % for math
\usepackage{amssymb}   % for math

\def\beq{\begin{equation}}
\def\eeq#1{\label{#1}\end{equation}}
\def\beqa{\begin{eqnarray}}
\def\eeqa#1{\label{#1}\end{eqnarray}}
\newcommand{\eeqn}{\end{equation}}

\newcommand{\leqn}[1]{(\ref{#1})}

\def\st{\tilde{t}}
\def\go{\tilde{g}}

\begin{document}
\title{
%%%%   Paper title goes here  %%%%%%%%%%%%%%
RPV SUSY with Same-Sign Dileptons at LHC-14} %% 
%***********************************************************************
% AUTHORS INFORMATION AREA
%***********************************************************************
\author{Michael Saelim and Maxim Perelstein
% Optional short acknowledgment: remove next line if non-needed
\thanks{This work is supported by the U.S. National Science Foundation through grant PHY-0757868 and CAREER award PHY-0844667.}
% DO NOT MODIFY THE FOLLOWING '\vspace' ARGUMENT
\vspace{.3cm}\\
% Addresses and institutions (remove "1- " in case of a single institution)
Laboratory for Elementary Particle Physics, Cornell University \\
Ithaca, NY 14853, USA
%% Remove the next three lines in case of a single institution
\vspace{.1cm}\\
}
%%***********************************************************************
% END OF AUTHORS INFORMATION AREA
%***********************************************************************

\maketitle

\begin{abstract}
We estimate the sensitivity of the 14 TeV LHC run to an R-parity violating supersymmetric model, via the same-sign dilepton (SSDL) signature. We consider the simplified model with light gluinos and stops, motivated by naturalness. We find that gluinos up to 1.4 TeV can be discovered with an integrated luminosity of 300 fb$^{-1}$. If a high-luminosity option is implemented and a 3000 fb$^{-1}$ dataset becomes available, the gluino mass reach can be increased to $1.6-1.75$ TeV.
\end{abstract}

Experiments at the 8 TeV run of the Large Hadron Collider (LHC-8) did not discover evidence for supersymmetry (SUSY), placing stringent lower bounds on the masses of many of the superpartners. In a completely natural theory, some of the superpartners are required to be in the few hundred GeV to a TeV range independently of the details of SUSY breaking. In particular, such model-independent upper mass bounds apply to stops $\st$ and gluinos $\go$ (see, for example, Ref.~\cite{Brust:2011tb}).
In the ``standard" realization of SUSY, with conserved R-parity and a weakly interacting lightest supersymmetric particle (LSP), the LHC experiments already strongly constrain this possibility. This motivates interest in alternative realizations of SUSY, including models with R-parity violation (RPV), which will be our focus in this contribution. On the theoretical side, an attractive scenario, ``MFV SUSY",  has been proposed~\cite{MFV_SUSY1,MFV_SUSY2}. In this model, RPV is confined almost exclusively to third-generation quarks, avoiding the bounds from the non-observation of baryon number violating processes. On the experimental side, RPV SUSY predicts events with no large missing transverse energy (MET), making it much more difficult to distinguish SUSY events from Standard Model (SM) backgrounds and thus avoiding most LHC constraints. 

Even though the standard large-MET cuts are not useful in RPV SUSY searches, other handles can be used. In this note, we will focus on the same-sign dilepton signature, proposed in~\cite{Baer:1996wa,AG,us,Durieux:2013uqa}. (For other signatures, see for example~\cite{Evans:2012bf,Han:2012cu,Duggan:2013yna,Bai:2013xla}.)  Consider a simplified RPV model with a stop LSP and a gluino in the TeV mass range, as motivated by naturalness. In simple SUSY models, the gluino is a Majorana particle, so that both decays
\beq
\go \to \st \bar{t},~~~\go \to \st^* t
\eeq{decays}
are possible and occur with the same probability. Gluino pair-production at the LHC can then lead to events with two {\it same-sign} tops, and if each of them decays leptonically, the same-sign dilepton (SSDL) signature is obtained. The stops decay via an RPV operator $\tilde{t}\to \bar{b}\bar{s}$, as predicted by the MFV SUSY scenario, so that no large MET is produced. (A small amount of MET is always present from neutrinos in leptonic top decays.) Nevertheless, this signature has a very small SM background, and provides a promising search channel. The 95\%~c.l. bound on the gluino mass, estimated in~\cite{us} based on a CMS search~\cite{CMS_SSDL} using $10.5$ fb$^{-1}$ of LHC-8 data, is about 800 GeV, approximately independent of the stop mass as long as the decay~\leqn{decays} can occur on-shell. Recently, this search strategy was implemented by the CMS collaboration in~\cite{CMS_new}, yielding the gluino mass bound of $900$ GeV with the full 2012 data set. The CMS analysis assumed that the decay~\leqn{decays} is off-shell. 

We estimate the sensitivity of the SSDL search for RPV SUSY at the next run of the LHC, for which we assume the center-of-mass energy of 14 TeV. To perform this estimate, we simulate signal events, for a set of 198 parameter points in the $(m_{\go}, m_{\st})$ plane, using {\tt PYTHIA 8}~\cite{pythia}. Signal cross sections are scaled up to NLO using K-factors calculated in {\tt Prospino 2.1}~\cite{prospino}. For gluino masses higher than what Prospino can calculate, we estimate the K-factors by linear extrapolation. We approximately model the detector response with {\tt Delphes 3}~\cite{delphes}, and apply an analysis based on the work in~\cite{us}. Pile-up effects are not included in this analysis.

The Delphes detector simulation is controlled by a modified version of the default CMS card included in Delphes. Particle propagation, tracking efficiencies, energy and momentum smearing, and calorimeter response are left unchanged. We require leptons to have $p_T > 20$ GeV and $|\eta| < 2.4$ (excluding electrons in the ECAL gap, $1.442 < |\eta| < 1.566$). We then apply an identification efficiency (73\% for $e$, 84\% for $\mu$) calculated in Appendix A of~\cite{us} from the data in~\cite{CMS_SSDL}, and an isolation cut requiring the scalar $p_T$ sum of all objects within $\Delta R = \sqrt{(\Delta \eta)^2 + (\Delta \phi)^2} = 0.3$ of the lepton to be no greater than 10\% of the lepton's $p_T$. We cluster fat jets with the anti-kt algorithm implemented in {\tt FastJet 3}~\cite{fastjet}, with $R = 1.0$, and require them to have $p_T > 40$ GeV and $|\eta| < 2.4$. For b-tagging, we use the tagging and mistagging efficiencies for the CSVM tagger from~\cite{btagging}. Finally, we also tag jets with masses above the top mass as ``high-mass jets''.

After the detector simulation, we select the same-sign lepton pair with the highest $p_T$ and pair invariant mass greater than 8 GeV to be our ``SSDL pair''. We apply a dilepton trigger efficiency of 96\% for $ee$, 93\% for $e\mu$, and 88\% for $\mu\mu$~\cite{CMS_SSDL}. We veto events where a third lepton, with $p_T > 10$ GeV and isolation sum no greater than 20\% of the lepton's $p_T$, forms an opposite-sign same-flavor pair with one of the SSDL pair leptons, with a pair invariant mass between 76 and 106 GeV, to remove background events with leptons from Z decays. We also veto events where a third lepton, with $p_T > 5$ GeV and isolation sum no greater than 20\% of the lepton's $p_T$, forms an opposite-sign same-flavor pair with one of the SSDL pair leptons, with a pair invariant mass below 12 GeV, to remove background events with leptons from a $\gamma^*$ or a low-mass bound state.

For an event to pass the analysis cuts, we require that an SSDL pair is found, and then we apply the event-level cuts for each of the signal regions used in the current CMS search, see Table 2 of Ref.~\cite{CMS_SSDL}, except for SR7. Additionally, we require either $N_{HMJ} \ge 1$ or $N_{HMJ} \ge 2$, where $N_{HMJ}$ is the number of high-mass jets in the event. To estimate the sensitivity of this search at LHC-14, we compare the projected reach for all 8 signal regions we implemented, with two $N_{HMJ}$ choices for each one. We select the most sensitive among these choices. However, we did not attempt to optimize the cuts further by going beyond the settings of the current CMS search; such optimization can of course only improve the reach. 

Irreducible backgrounds to this search involve processes that can generate two prompt same-sign leptons. To model them, we generate 500K events each of $t\overline{t}W^\pm$ and $t\overline{t}Z$ in {\tt MadGraph 5}~\cite{madgraph}, and scale their cross sections up to NLO with K-factors 1.236~\cite{kfactorttw} and 1.387~\cite{kfactorttz}, respectively.  These events are put through the same hadronization, detector simulation, and analysis as the signal events.

Instrumental backgrounds to this search mainly involve ``fake leptons'' from heavy-flavor decays and misidentified hadrons, and ``charge flips'' where an opposite-sign dilepton event has one of its lepton charges mismeasured. These dominantly arise from top pair-production. Since both effects are rare, a very large sample of $t\bar{t}$ Monte Carlo events would be required to estimate their rates from simulation. In this preliminary study, we instead estimate the instrumental background by noting the rates in each of the signal regions in Table 2 of~\cite{CMS_SSDL} and assuming that they scale with the $t\overline{t}$ cross section when the collision energy is increased from 8 to 14 TeV. Previous study in~\cite{us} showed that the resulting reach is not heavily sensitive to this assumption. We combine the irreducible and instrumental backgrounds to arrive at the total background for each signal region and choice of $N_{HMJ}$ cut. 

\begin{figure}
\begin{center}
\includegraphics[width=3in]{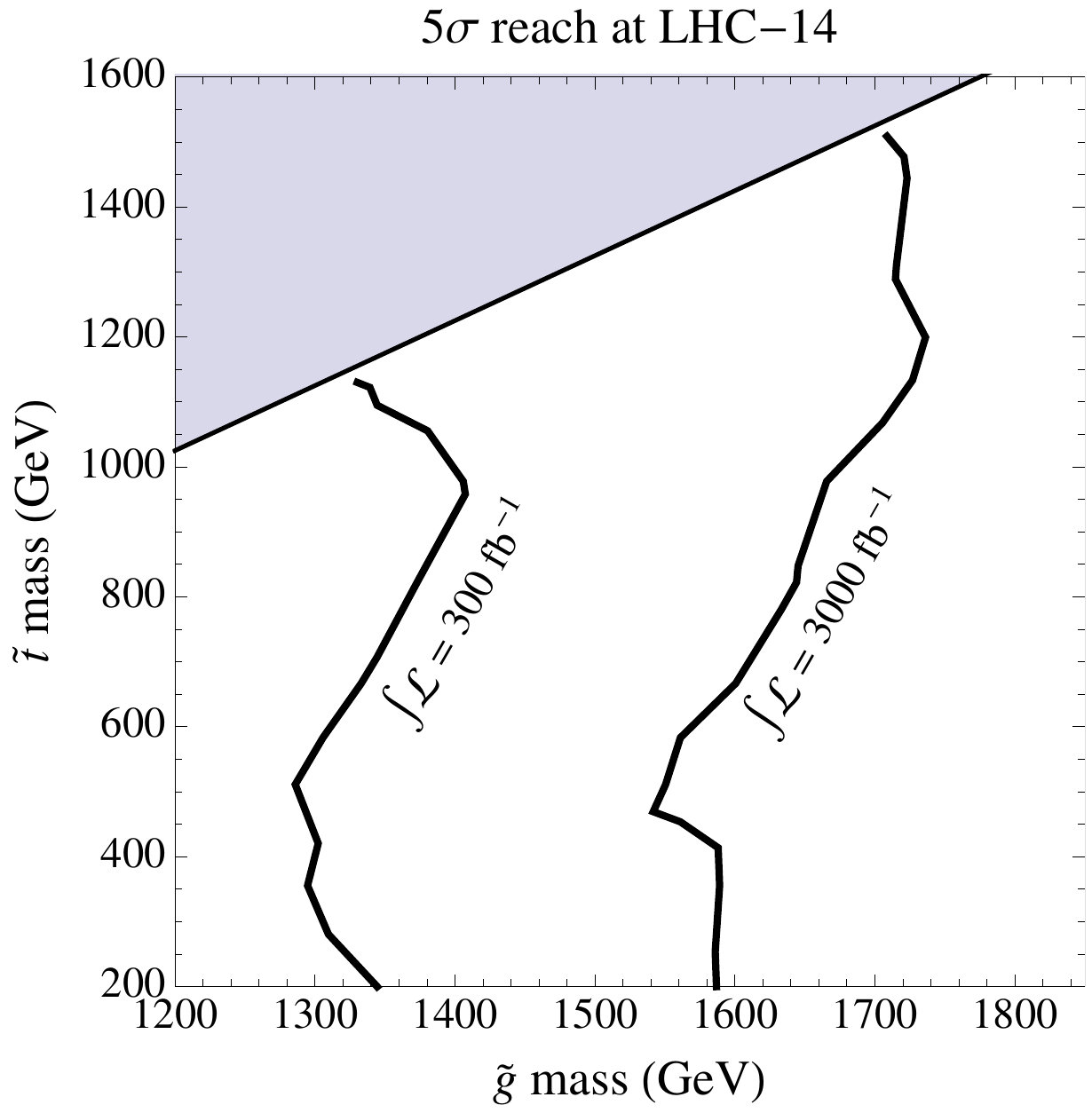}
\end{center}
\caption{Estimated 5$\sigma$ reach of LHC-14 in the RPV SUSY simplified model.}
\label{fig:reach}
\end{figure}

The projected sensitivity of the LHC-14 in the $(m_{\go}, m_{\st})$ plane is shown in Figure~\ref{fig:reach}. We find that among the CMS signal regions, SR5, with the additional requirement of 2 high-mass jets, gives the best sensitivity, although several other SRs are almost as sensitive. Significant improvements of the reach compared to the present exclusion bounds can be clearly achieved: for example, with 300 fb$^{-1}$ of data, the $5\sigma$ reach in gluino mass is as high as $1.4$ TeV, approximately independent of $m_{\st}$. 
If the High Luminosity (HL) LHC upgrade is implemented and yields 3 ab$^{-1}$ of data, the reach can be further increased to $1.6-1.75$ TeV depending on $m_{\st}$. Note that for technical reasons, our analysis has been restricted to the part of parameter space where the decays~\leqn{decays} occur on-shell. However, the analysis should retain sensitivity even in the region where the gluinos decay in a single-step, three-body channel $\go\to tbs$.  

In summary, same-sign dilepton (SSDL) signature provides a promising channel to search for RPV supersymmetry, within the gluino/stop simplified model motivated by naturalness. We find that experiments at the LHC-14 can achieve impressive sensitivity for this model, especially if a large data set becomes available with the HL upgrade. 

\begin{footnotesize}
% IF YOU DO NOT USE BIBTEX, USE THE FOLLOWING SAMPLE SCHEME FOR THE REFERENCES
% ----------------------------------------------------------------------------

% ----------------------------------------------------------------------------

\end{footnotesize}

% ****************************************************************************
% END OF BIBLIOGRAPHY AREA
% ****************************************************************************

\end{document}